Abstract

Applying the concept of triadic closure to coauthorship networks means that scholars are likely to publish a joint paper if they have previously coauthored with the same people. Prior research has identified moderate to high (20% to 40%) closure rates; suggesting that this mechanism is a reasonable explanation for tie formation between future coauthors. We show how calculating triadic closure based on prior operationalizations of closure, namely Newman's measure for one-mode networks (NCC) and Opsahl's measure for two-mode networks (OCC), may lead to higher amounts of closure as compared to measuring closure over time via a metric that we introduce and test in this paper. Based on empirical experiments using four large-scale, longitudinal datasets, we find a lower bound of about 1~3% closure rates and an upper bound of about 4~7%. These results motivate research on new explanatory factors for the formation of coauthorship links.

1. Introduction and Background

In the field of collaboration networks research, scholars aim to understand the mechanisms that underlie the formation of relationships between authors, which result in coauthorship ties and networks. One of these mechanisms is homophily (McPherson, Smith-Lovin, & Cook, 2001), i.e., the tendency of scholars to be more likely to collaborate with each other if they share personal, behavioral, sociodemographic or other attributes. Prior work on this topic has shown that sociologists in three European countries had a tendency to work together if they were of the same gender, belonged to the same department, or had similar citation patterns (Hâncean & Perc, 2016; Hâncean, Perc, & Vlăsceanu, 2014). Homophily-based tie formation in collaboration networks was also confirmed in a study of collaboration patterns among Turkish engineering scholars over a period of 33 years, where it was found that authors are likely to publish together if they had similar degrees, which could be a proxy for similar backgrounds, and were of similar academic ages (Türker & Çavuşoğlu, 2016).

Unlike the aforementioned studies that mostly leverage similarities in node-level features, another stream of collaboration networks research focused on the patterns in network structure in order to model the formation of scientific collaboration networks. This approach relates to the long-held observation that people have a natural tendency to engage in small, tightly knit groups, even though coauthorship networks might differ from non-professional types of relationships, such as friendship (Holland & Leinhardt, 1970). This tendency has been studied under the term 'clustering' or 'community detection' in the networks literature, and several algorithms and metrics have been proposed to detect and measure this effect. For example, Watts and Strogatz (1998) introduced the clustering coefficient, which measures the density of a node's ego network. Barabási et al. (2002) interpreted this clustering in collaboration networks as the likelihood of an author's collaborators to work together, and referred to the metric as the local clustering coefficient.

Newman (2001b) extended this notion to the graph level by proposing a new measure, i.e. the Global Clustering Coefficient, and used this measure to describe link formation mechanisms in collaboration networks. Comparable to the original notion of triadic closure (Rapoport, 1953), Newman's global clustering coefficient is concerned with the clustering of trios (groups of three nodes): if node A is connected to nodes B and C, but B and C are not yet linked, then B and C are likely to form a link in the future. Although Newman (2001b) referred to this tendency as transitivity, it is better known in the social network literature as triadic closure (Kossinets & Watts, 2006). In our paper hereafter, triadic closure is used synonymously with Newman's clustering tendency (what he called transitivity)[1]. In order to measure triadic closure on the network level, Newman (2001b) just proposed a clustering coefficient to calculate the ratio of the number of trios that are triads (i.e., three nodes connected by three edges) over the number of three nodes that are connected by two edges. Applying this measure to scientific collaboration networks, he found that triadic closure happens more often in computer science and physics than in biomedicine.

Newman's global clustering coefficient (NCC) has been adopted as a proxy for the likelihood of two scholars who share a common third coauthor, but who have not yet published together, to collaborate with each other on a joint paper in the future. For example, Newman (2001b) found a clustering coefficient of 0.43 for a coauthorship network of 52,909 authors from 98,502 papers available from the Los Alamos e-Print Archive. Based on this result, he argued that "two scientists typically have a 30% or greater probability of collaborating if both have collaborated with a third scientist" (Newman, 2001b, p.408). Until recently, scholars have often followed Newman's way of interpreting clustering coefficients for triadic closure in coauthorship networks. Table 1 summarizes selected empirical results for measuring triadic closure based on NCC at the field level (e.g., Franceschet, 2011; Kim & Diesner, 2015), institutional level (e.g., Türker, Durgut, & Çavuşoğlu, 2016), and national level (e.g., Çavuşoğlu & Türker, 2013; Kim, Tao, Lee, & Diesner, 2016; Perc, 2010).

---

[1] Transitivity seems to be associated more commonly with directed networks (Wasserman & Faust, 1994) than with undirected ones. Many network research papers and software packages implementing the Newman metric (2001b) still use transitivity to refer to triadic closure.

*Table 1: Selected Studies Reporting Clustering Coefficients for Triadic Closure. Eight papers reported clustering coefficients for collaboration networks from six fields – Computer Science, Math, Medicine, Neuroscience, Physics, and Sociology - and three nations – Korea, Slovenia, and Turkey. The coefficients range from 0.066 (Medicine) to 0.76 (Neuroscience). For Mathematics and Neuroscience in Barabási et al. (2002), the coefficients were estimated from Figure 4 using an open source image analysis tool, ImageJ (https://imagej.nih.gov/ij/)*

| Field or Country | Years | No. of Papers (No. of Authors) | Avg. authors per paper | Clustering Coefficient | Reference |
|---|---|---|---|---|---|
| Computer Science | 1936-2008 | 1,216,526 (731,333) | 2.56 | 0.24 | Franceschet (2011) |
| Computer Science | 1995-1999 | 13,169 (11,994) | 2.22 | 0.496 | Newman (2001b) |
| Korea | 1948-2011 | 703,073 (415,695) | 2.79 | 0.19 | Kim et al. (2016) |
| Mathematics | 1940-1999 | 1,598,000 (337,000) | 1.45 | 0.15 | Grossman (2002) |
| Mathematics | 1991-1998 | 70,901 (70,975) | - | 0.59 | Barabási et al. (2002) |
| Medicine | 1995-1999 | 2,163,923 (1,520,251) | 3.75 | 0.066 | Newman (2001b) |
| Neuroscience | 1991-1998 | 210,750 (209,293) | - | 0.76 | Barabási et al. (2002) |
| Physics | 1995-1999 | 98,502 (52,909) | 2.53 | 0.43 | Newman (2001b) |
| Physics | 1995-1999 | 66,652 (56,627) | 8.96 | 0.726 | Newman (2001b) |
| Slovenia | 1960-2010 | 76,194 (7,380) | - | 0.20 | Perc (2010) |
| Sociology | 1963-1999 | 281,090 (197,976) | 1.56 | 0.19 | Moody (2004) |
| Turkey | 1980-2010 | 237,409 (151,745) | 4.08 | 0.75 | Çavuşoğlu and Türker (2013) |

One problem with this common procedure is that the NCC may report false positives for triadic closure, especially in scientific collaboration networks, as illustrated with an example in Table 2. In a collaboration network, authors are nodes that get connected if they appear in the same byline of a publication. In Case 1, authors Y and Z collaborate with author X on papers A and B, respectively, and thus form edges with author X (i.e., X-Y and X-Z). These two pairs of edges are called a 2-path. In paper C, authors Y and Z collaborate, which completes a triangle of edges among author X, Y, and Z, and this is exactly the effect that triadic closure is supposed to represent. Here, the 2-path is said to be closed by the edge Y-Z[2]. The NCC counts the number of closed 2-paths over all 2-paths in a network. In contrast, in Case 2, author X, Y, and Z are fully connected to each other, seemingly leading to triadic closure because these authors appear in the same byline of paper D. However, the NCC does not distinguish these two cases and may thus inflate the tendency of triadic closure in a coauthorship network, especially in data where papers written by three or more people are common (Opsahl, 2013).

*Table 2: Visualization of the Concept of Triadic Closure according to Newman's Clustering Coefficient (NCC). Circles represent authors, while squares represent papers. CASE 1 shows the typical situation of triadic closure where authors Y and Z who have once collaborated with X on different papers start to work with each other later. NCC captures this well. In CASE 2, however, NCC falsely reports that triadic closure happens.*

| Case | Network Visualization |
|---|---|
| CASE 1)<br>Paper A: authors X, Y<br>Paper B: authors X, Z<br>Paper C: authors Y, Z | 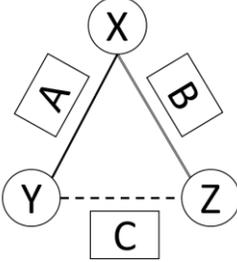 |
| CASE 2)<br>Paper D: authors X, Y and Z | 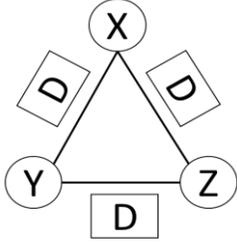 |

Technically, this problem arises mainly when Newman's measure is directly applied to one-mode networks projected from a two-mode (i.e., bipartite) network. A two-mode network has two types of nodes (e.g., papers and authors), and edges only across nodes of different types. A two-mode network can be projected onto two one-mode networks. For example, a two-mode network of papers and authors can be transformed into one network of papers where two papers (nodes) are connected if they are written by the same authors. Thus, a coauthorship network is a special case of a one-mode projection from a two-mode network (of papers and authors) where people are linked because they coauthored a paper. As shown in Case 2 of Table 2, if three or more authors who jointly wrote a paper as per the two-mode network, the one-mode network of authors will have triangles solely as a result of the projection process, not because of triadic closure.

---

[2] The triadic closure contains three cases of 2-path closure: (1) Y-X-Z closed by Y-Z, (2) X-Z-Y closed by Y-X, and (3) Z-Y-X closed by X-Z. This also applies to Case 2 in Table 2.

To overcome the aforementioned limitation, several scholars have proposed alternative clustering measures (for examples see Lind, Gonzalez, & Herrmann, 2005; Opsahl, 2013; Robins & Alexander, 2004), including Newman himself (Newman, Strogatz, & Watts, 2001). Despite their variations, the common theme with these alternatives is that clustering should be calculated directly on two-mode networks in order to exclude artifactual triangles as in Case 2 above. Among them, the Opsahl (2013) measure is designed to capture triadic closure in two-mode scientific collaboration networks as illustrated in Table 3 ((Opsahl, 2013) p. 162). Opsahl (2013) defines triadic closure in a two-mode network as a closed 4-path (as opposed to Newman (2001b), who defines triadic closure as closed 2-paths in one-mode networks). For example, in Table 3, Case 1, a 4-path (Y-A-X-B-Z) is closed by Y-C-Z[3]. Opsahl (2013) applied this measure (hereafter referred to as OCC), and found a triadic closure (clustering coefficient) of 0.28 for a coauthorship network of 16,726 physicists and 22,016 papers, which is considerably lower than when NCC is applied to the same data (clustering coefficient: 0.36)[4].

*Table 3: Visualization of the Concept of Triadic Closure according to Opsahl's Clustering Coefficient (OCC). Circles represent authors, while squares represent papers. CASE 1 shows the typical situation of triadic closure where authors Y and Z, who have previously collaborated with X on different papers, work with each other later. OCC captures this effect through a two-mode network approach to triadic closure. Unlike NCC, OCC does not report false triadic closure in CASE 2.*

| Case | Network Visualization |
|---|---|
| CASE 1) <br> Paper A: author X and author Y <br> Paper B: author X and author Z <br> Paper C: author Y and author Z | 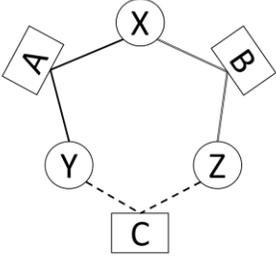 |
| CASE 2) <br> Paper D: author X, author Y, and author Z | 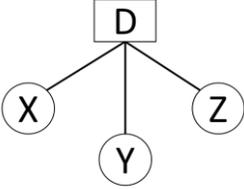 |

---

[3] The triadic closure contains three cases of 4-path closure: (1) Y-A-X-B-Z closed by Y-C-Z, (2) X-B-Z-C-Y closed by Y-A-X, and (3) Z-C-Y-A-X closed by X-B-Z

[4] Although calculated against the same dataset, the clustering coefficient by the Newman (2001b) method in Opsahl (2013) is 0.3596, while the one in Newman (2001b) is 0.348.

Opsahl (2013)'s measure can exclude false positives in terms of triadic closure, which are generated by the projection of two-mode networks onto one-mode networks, but another problem, which also applies to the NCC, remains unsolved. Practically speaking, actual triadic closure in scientific collaboration networks entails a time lag between (1) the formation of 2-paths in a one-mode network as well as 4-paths in two-mode networks, and (2) their closure. This can happen in an asynchronous or simultaneous manner. For instance, in the example of Case 1 in Table 2 and Table 3, author X can collaborate with Y first, and then with Z later (asynchronous), or with Y and Z on two different papers simultaneously. However, the new collaboration between Y and Z should happen after both the collaborations between X and Y and between X and Z have occurred. This seems logical, especially if we want to use triadic closure as a proxy for the probability of two scholars forming a new coauthoring tie conditioned on their shared collaborator(s) as conceptualized by Newman (2001b) and his followers[5]. The clustering measurements proposed by Newman (2001b) and Opsahl (2013) are both calculated on a static snapshot of a network and, therefore, may include the clustering of three nodes that are not an instance of triadic closure but wrongly counted as such. In Case 1 of Table 3, for example, paper C might be written before papers A and B were produced. In other words, authors Y and Z may have collaborated with each other before each of them worked with author X. Consequently, the collaboration between Y and Z is not related to their shared coauthor X. According to Newman and Opsahl's measures, however, this case is counted as an instance of triadic closure.

Therefore, we argue that triadic closure in collaboration networks has not been appropriately measured because the temporal aspect of this process was disregarded. Both of the aforementioned measures, whether they are calculated on one-mode or two-mode networks, use a snapshot of three nodes being fully connected, ignoring which edges formed first. Subsequently, scholars have used the static approach to infer a specific edge formation mechanism that inevitably requires time to be considered. For example, based on clustering coefficients obtained via the Newman (2001b) measure, many coauthorship network studies have concluded that scholars show a strong tendency to close links to shared coauthors (see Table 1). Except for the field of biomedicine (0.066), the average clustering coefficient of eight sampled collaboration networks is 0.40, indicating that triadic closure happens in nearly 40 percent of the studied cases. A part of these results might be cases of artifactual triadic closure due to network projection or the reversed temporal order of closure. In this paper, we address this issue by empirically identifying how often triadic closure truly happens instead of just estimating it. This is done by considering the temporal order of open 4-paths and their closing in two-mode networks. We compare our results to those obtained by applying Newman's and Opsahl's metrics to the same large-scale collaboration networks from three fields (computer science, physics, and biomedicine) and one nation (South Korea) over a time period of almost 20 years. Using this strategy allows us to uncovers the actual number of triadic closures in scientific collaboration networks. In the next section, we describe the datasets and define three different measures for triadic closure.

2. Methodology
2.1. Datasets

The following four coauthorship datasets are analyzed: MEDLINE (biomedicine), DBLP (computer science and informatics), APS (Physics), and KISTI (country-level data for Korea). Table 4 summarizes the characteristics of these datasets. For each dataset, papers published between 1991 (or 1996) and 2009 were considered. In the original MEDLINE and APS datasets, author names are not disambiguated, while the DBLP and KISTI datasets contain disambiguated author names. Prior research has shown that insufficient attention to author name disambiguation can strongly bias network properties including clustering coefficients (Fegley & Torvik, 2013; Kim & Diesner, 2016). In order to avoid this effect, we obtained an algorithmically disambiguated version of the MEDLINE and APS datasets (details below).

---

[5]Opsahl (2013) never uses the clustering coefficient defined for two-mode networks as an indicator of the probability of two scientists collaborating when they have a third coauthor in common.

*Table 4: Summary of Datasets. Four datasets representing three fields and one nation were used for analysis. Each dataset was carefully filtered from its original source such that size and time coverage are comparable.*

| Dataset | Field | Period of Analysis | Number of Papers[6] | Avg. Number of Authors per Paper[7] |
|---|---|---|---|---|
| DBLP | Computer and Information Science | 1991-2009 | 231,161 | 2.91 |
| APS | Physics | 1991-2009 | 241,329 | 3.80 |
| MEDLINE | Biomedicine | 1996-2009 | 302,293 | 4.91 |
| KISTI | Domestic Publication in Korea | 1991-2009 | 273,869 | 3.17 |

2.1.1. DBLP

The Digital Bibliography & Library Project (DBLP) contains publication records from computer and information science starting in the 1950s. In DBLP[8], author names are disambiguated by (1) matching author name strings, (2) calculating coauthor similarity based on coauthor matching of authors and their coauthor's coauthors, and (3) manual correction of errors that were reported by scholars (Reitz & Hoffmann, 2011). According to Kim and Diesner (2015), disambiguation accuracy in DBLP is reported to be 0.952 (K-metric) and 0.96 (Pairwise F1). DBLP indexes more than 2 million papers published in journals and conferences. For this study, over 231,000 papers published in 392 computer science journals[9] were used as a sample of DBLP data.

2.1.2. APS

The American Physical Society (APS) maintains publication records from the Physical Review journals dating back to 1893. Author names in the original APS dataset[10] are not disambiguated. In order to correct for that, we applied the name disambiguation algorithm described in Martin, Ball, Karrer, and Newman (2013), which utilizes name string, coauthor name, affiliation, and venue information to determine the match of a pair of author names. The algorithm's accuracy is reported to merge 3% of sampled authors incorrectly and to split 12% of sampled authors incorrectly (Martin et al., 2013). Out of approximately 540,000 papers, 241,329 papers spanning 1991-2009 were selected for this study.

---

[6] In each dataset, the distribution of the number of authors per paper was calculated. Then, we identified a threshold value for the number of authors per paper such that 98%~99% of the papers in our data are written by that many or less authors: DBLP (7), APS (14), MEDLINE (14), and KISTI (8). Papers with a higher number of authors were excluded from analysis.
[7] Single-authored papers are excluded from analysis.
[8] http://dblp.uni-trier.de/xml/
[9] The list of 392 journals was obtained from Thomson Reuters Journal Citation Report 2012 for the Computer Science category. Then, those journals' names and papers published in these journals were searched for in DBLP.
[10] http://journals.aps.org/datasets

### 2.1.3. MEDLINE

The National Library of Medicine's (MEDLINE)[11] bibliographic database indexes journal papers from biology and medicine published from 1950 to the present. Each paper is recorded with a unique identifier (PMID), author name(s), author affiliation (if available), title, publication venue, and keywords (also referred to as medical subject headings (MeSH)). We obtained the data through the Author-ity database (Torvik & Smalheiser, 2009), where author names in MEDLINE are algorithmically disambiguated based on similarity calculation of the author name, coauthor name, title word, journal, and the MeSH. The accuracy of these data is reported to be up to 98~99%. We only considered terms with the MeSH term 'brain' since this is one of the most frequent MeSH terms in MEDLINE. This filtering resulted in approximately 302,000 papers out of about 10 million papers published between 1991 and 2009.

### 2.1.4. KISTI

The Korea Institute of Science and Technology Information (KISTI) dataset[12] includes around 710,000 publication records from both conference proceedings and journals published in Korea beginning in the late 1940s until 2016. In the KISTI data, author names are disambiguated by a clustering algorithm utilizing features like name string, affiliation, coauthor name, title, and publication venue. The accuracy is reported to be 0.94 in terms of pairwise F1. After the computational disambiguation, people at KISTI performed manual inspection for the remaining cases that were unclear. Since previous studies mostly measured triadic closure for journal papers, we only considered 273,869 journal papers in this study for consistency.

## 2.2. Measurements

To measure how often triadic closures truly occur in scientific collaboration networks, we used three different measurements. First, the Newman (2001b) Clustering Coefficient (NCC), which calculates the ratio of the number of closed 2-paths over the number of 2-paths in a network as in Equation (1). Since the NCC is calculated on one-mode network data (i.e., an author-by-author matrix), the byline information (i.e., author list per paper) in each dataset was converted into an edge list to produce an author-by-author network matrix, where two author names are connected if they appear in the same byline. the NCC was then calculated by using R package *igraph* (Csardi & Nepusz, 2006).

$$NCC = \frac{Number\ of\ closed\ 2\ paths}{Number\ of\ 2\ paths} \quad (1)$$

Second, the Opsahl (2013) Clustering Coefficient (OCC) calculates the ratio of the number of closed 4-paths over the number of 4-paths in a network as in Equation (2). Since the OCC is defined for two-mode networks, the byline information for each dataset was transformed into an edge list, where an author name is connected to any paper (co-)authored by this person. The OCC was calculated with the R package *tnet* (Opsahl, 2009).

---

[11] https://www.nlm.nih.gov/bsd/licensee/medpmmenu.html
[12] http://scholar.ndsl.kr/index.do

$$OCC = \frac{Number\ of\ closed\ 4\ paths}{Number\ of\ 4\ paths} \quad (2)$$

Third, a measure that we introduce with this paper, namely the Over-Time Clustering Coefficient (TCC), extends the Opsahl (2013) approach by dividing the number of node pairs embedded in closed 4-paths in a network at time '$x + 1$' by the number of node pairs embedded in 4-paths in a network at time '$x$.'

$$TCC = \frac{Number\ of\ node\ pairs\ embedded\ in\ closed\ 4\ paths\ at\ t_{x+1}}{Number\ of\ node\ pairs\ embedded\ in\ 4\ paths\ at\ t_x} \quad (3)$$

The first difference between OCC and TCC lies in the treatment of multiple 4-paths involving two nodes. In Table 5, Case 2 shows an example where a paper (C) and an author (W) are added to Case 1. According to the OCC, this addition increases the number of 4-paths in the network: i.e., in Case 1, a single 4-path exists, while in Case 2, three 4-paths are found[13]. Although Case 1 and Case 2 have a different number of 4-paths, these two cases can be considered as the same situation where the closure between nodes Y and Z is of major interest. Based on this proposition, the TCC focuses on whether two target nodes are embedded in any open 4-path while ignoring how many 4-paths involve these two nodes. Thus, for the TCC, Case 1 and Case 2 are the same situation. For example, if nodes Y and Z are connected in each case, the TCC produces the same value (=1.0).

*Table 5: Visualized Concept of Triadic Closure by Over-Time Measure (TCC). Circles represent authors, squares represent papers. CASE 1 is a typical situation where authors Y and Z are embedded in a 4-path, which NCC and TCC count the same way. In CASE 2, however, OCC counts three 4-paths for Y and Z, while TCC counts a single 4-path for them because TCC considers only the embeddedness of Y and Z in any 4-path ignoring how often they appear together in 4-paths.*

| Case | Network Visualization |
|---|---|
| CASE 1<br>Paper A: author X and author Y<br>Paper B: author X and author Z | (network diagram with nodes A, B, X, Y, Z) |
| CASE 2<br>Paper A: author W, author X, and author Y<br>Paper B: author W, author X, and author Z<br>Paper C: author X and author Y | (network diagram with nodes W, C, A, X, B, Y, Z) |

---

[13] The 4-path in Case 1 is Y-A-X-B-Z. The 4-paths in Case 2 are: (1) Y-A-X-B-Z, (2) Y-A-W-B-Z, and (3) Y-C-X-B-Z.

The second difference between the OCC and the TCC is the consideration of time: When it comes to time slicing, coauthorship networks often are partitioned on an annual basis, and data are compared from time *t* to time *t+1* (Barabási et al., 2002; Liben-Nowell & Kleinberg, 2007; Newman, 2001a). In our study, we define a target year ($t_{x+1}$) as a one-year period and its prior year ($t_x$) as a period of five years. For example, if the target year is 2009, the prior year is the period from 2004 to 2008. Then, we identify authors who appear both in the target year (e.g., 2009) and its preceding five years (e.g., 2004-2008). Next, we retrieve all author pairs embedded in 4-paths and that were not in closed 4-paths during the five preceding years. Finally, we calculate how many of those node pairs embedded in 4-paths also are embedded in closed 4-paths in the target year. The values can range from zero (none of author pairs are embedded in closed 4-paths) to one (all pairs are embedded in closed 4-paths).

3. Results

Conceptualizing and calculating triadic closure in different ways leads to different results. Figure 1 illustrates the over-time change in the clustering coefficient, i.e. the ratio of triadic closure, based on the Newman's Clustering Coefficient (NCC) (circles), the Opsahl's Clustering Coefficient (OCC) (squares), and the Over-Time Clustering Coefficient (TCC) (triangles). In this figure, each time point on the *x*-axis represents the last five years for the NCC and OCC. For example, the NCC and OCC for the year 2009 are calculated against the DBLP coauthorship network covering 2005-2009 (5 years). For the TCC, each year in the chart refers to a target year in which the closure of 4-path happens, and the preceding five years are used to detect open 4-paths.

Depending on the operationalization of the clustering coefficient, different results are obtained across all datasets and periods. For example, in the DBLP-based coauthorship networks, the average NCC of all 5-year windows is 0.429, the OCC is 0.152, and the TCC is 0.031. First, the differences between the NCC and OCC are not unexpected: the NCC does not distinguish artificial triadic closure introduced by the projection from two-mode to one-mode networks.

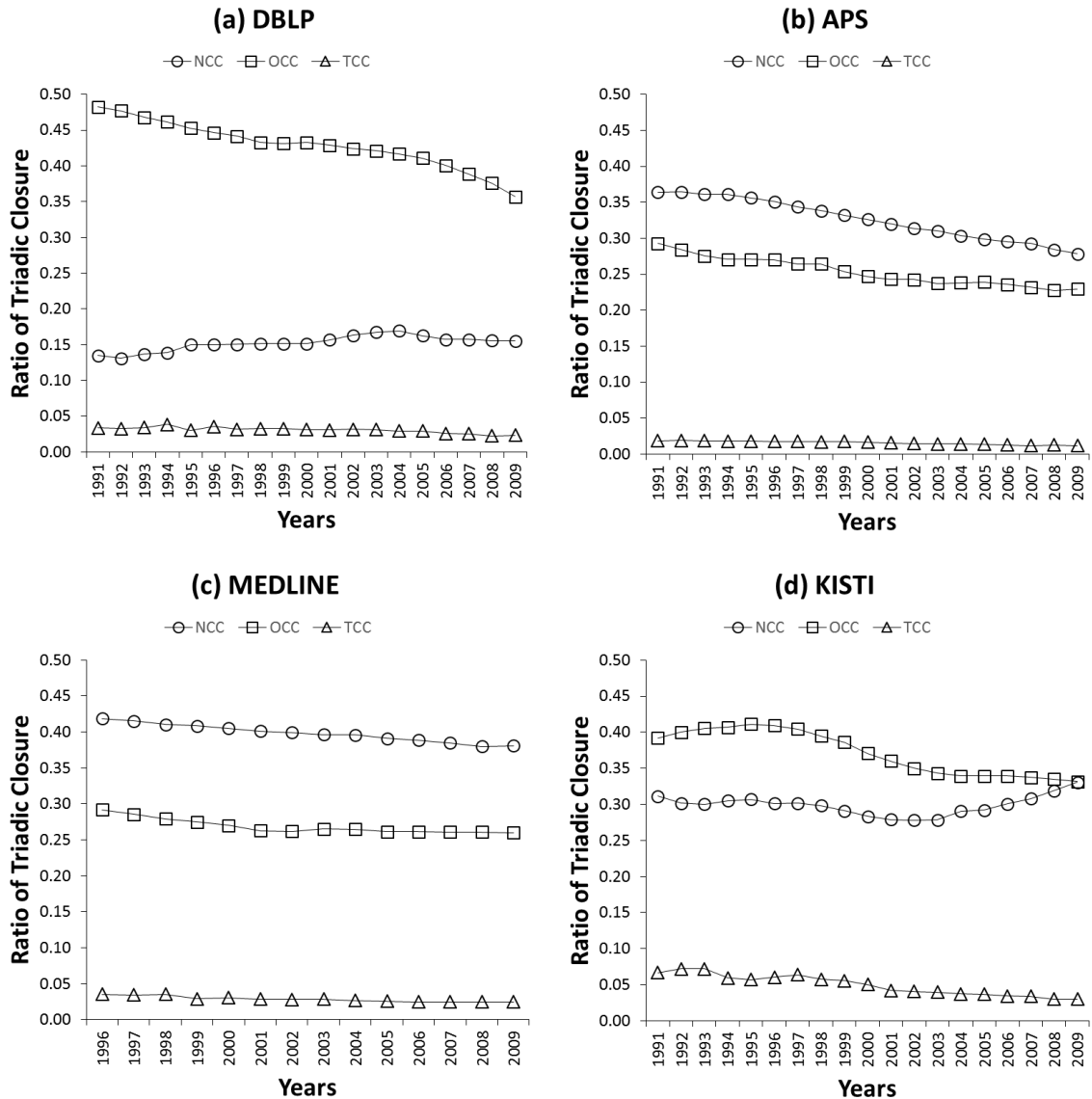

*Figure 1: Triadic Closure in Four Datasets Measured via Three Different Measurements. NCC = Newman's (2001b) Clustering Coefficient, OCC = Opsahl's (2013) Clustering Coefficient, and TCC = Over-Time Clustering Coefficient. NCC and OCC measure triadic closure for a five-year period (e.g., 2001-2005), which is denoted on the x-axis by the last year of the period (e.g., 2005). TCC counts how many 4-paths in any preceding four years (e.g., 2001-2004) are closed in a target year (e.g., 2005). In all datasets and over time, measuring actual triadic closure via TCC leads to considerably lower values than when measuring those values via NCC and OCC.*

Triadic closure, when measured based on the NCC and TCC, decreases over time, but increases slightly according to the OCC, especially for the DBLP and KISTI data. An explanation for the change in the OCC value might be that OCC detects triadic closure that the NCC cannot capture.

Another finding is that the values for the NCC and OCC become closer to each other when considering long time periods. For example, using the (d) KISTI data, the values for the NCC and OCC differ considerably in the beginning, but start to converge around the late 2000s[14]. Table 6 provides an explanation for this effect: there, the three measures produce different values when being applied to the same case. First, according to the OCC, a total of seven 4-paths are found, with five of them being closed (OCC = 4/7 = 0.71)[15]. In contrast, the NCC-based approach results in five 2-paths, and three of them are closed (NCC = 3/5 = 0.60)[16]. The difference in values between the OCC and NCC comes from the treatment of papers A and B. According to the NCC, multiple collaborations between X and Y (i.e., they coauthored A and B) are ignored (i.e., counted only once). The OCC, however, counts each collaboration between X and Y separately: X and Y are embedded in two different 2-paths (i.e., Y-B-X and Y-A-X). When multiple collaborations between authors are common in a network, the OCC will result in higher scores than the NCC. The TCC provides a different value from the NCC and OCC for the same case: only one node pair is embedded in closed 4-path (TCC = 1/2 = 0.50)[17] when we assume that paper E is written in a target year and other papers A, B, C, and D are written in preceding years.

*Table 6: Visualized Operationalization of Triadic Closure by NCC, OCC and TCC. Repeated collaboration between two authors is counted only once by NCC. In contrast, OCC considers each collaboration instance as unique for counting 4-paths, which can lead to a higher count of triadic closure than NCC. TCC only cares about embeddedness of two authors in any 4-path, ignoring repeated collaborations between them.*

| Case | Measure | Network Visualization |
|---|---|---|
| Paper A: authored by X, Y<br>Paper B: authored by X, Y<br>Paper C: authored by X, Z<br>Paper D: authored by W, Z<br>Paper E: authored by Y, Z | OCC<br>TCC | (network diagram with nodes B, A, X, C, Y, Z, D, E, W) |
| | NCC | (network diagram with nodes X, Y, Z, W) |

The most noticeable finding from this work is that, across all datasets and over time, measuring actual closure via TCC leads to considerably lower values than when calculating those values based on the NCC and OCC. For example, in the DBLP-based coauthorship networks, on average, 42.9% of the 2-paths are closed according to the NCC, 15.2% of the 4-paths are closed when using the OCC, and only 3.1% of the 4-paths are closed based on the TCC. This pattern is similar for the other datasets; TCC leads to average closure values of 1.6% in ASP, 2.9% in MEDLINE, and 4.9% in KISTI, while NCC and OCC result in 25% and higher closure ratios.

---

[14] In 2010, OCC (0.34) surpasses NCC (0.33).
[15] 4-paths by OCC: (1) Y-B-X-C-Z (closed by Z-D-Y), (2) Y-A-X-C-Z (closed by Z-D-Y), (3) X-C-Z-D-Y (closed by Y-A-X or Y-B-X), (4) Z-D-Y-A-X (closed by X-C-Z), (5) Z-D-Y-B-X (closed by X-C-Z), (6) X-C-Z-E-W, and (7) W-E-Z-D-Y
[16] 2-paths by NCC: (1) Y-X-Z (Y-Z), (2) X-Z-Y (Y-Z), (3) Z-Y-X (X-Z), (4) W-Z-X, and (5) W-Z-Y
[17] 4-paths by TCC: (1) Y-A & B-X-C-Z (closed by Z-D-Y) and (2) X-C-Z-E-W

One problem with the TCC is that it does not consider all open 4-paths within the 5 years preceding the target year. This is because the TCC is applicable only to nodes (i.e., authors) that appear both in a target year and the preceding 5 years. We tested whether this assumption is reasonable as follows: Table 7 shows how many nodes from a target year also have appeared in the preceding five years for each dataset. In DBLP, for example, nodes that appear both in the target year and its preceding years constitute on average 45.92% of all nodes in target years and on average 18.50% of all nodes in preceding years. In order to investigate how the number of preceding years impacts TCC, we varied this parameter (1 to 5 years). Figure 2 shows the ratio of triadic closure measured via TCC depending on the number of preceding years. We find that the amount of triadic closure decreases when the number of preceding years is increased. For example, in each dataset, considering one preceding year yields almost two times larger ratios of clustering coefficients than using five preceding years. One possible explanation of this effect might be that as the number of preceding years increases, more node pairs embedded in open 4-paths (denominator) can be found. However, not all of these pairs (numerator) are embedded in closed 4-paths in the target year. Another reason may be that some triadic closures are not captured by the TCC, especially with lengthy windows into the past. This may happen when, if we set the value for preceding years to five, for example, triadic closure that takes place during the five years are excluded from the TCC calculation.

The results also show that the ratio of triadic closure tends to decrease over time in all dataset and regardless of the number of preceding years. This tendency may be a function of network size and/ or a result of changes in the underlying personal, professional, organizational or other social (dis)incentives and mechanisms for tie formation among collaborators. Further research is needed to identify and distinguish these possible reasons.

Table 7: Average Ratio of Number of Nodes Appearing Both in a Target Year and Preceding Years Against Total Number of Nodes in a Target Year and Preceding Years. Standard deviation is shown in percentage points in parentheses. On average, 45.92%~69.94% of authors who appeared in target years also appeared in its preceding five years across four datasets. In contrast, 17.42%~31.23% of authors who appeared in the preceding five years also appeared in the target years.

| Datasets | DBLP | APS | MEDLINE | KISTI |
|---|---|---|---|---|
| Target Year[18] | 45.92% (3.14%$p$) | 69.94% (2.31%$p$) | 50.96% (1.64%$p$) | 52.75% (3.53%$p$) |
| Preceding Years | 18.50% (1.32%$p$) | 31.23% (1.13%$p$) | 17.42% (1.05%$p$) | 23.35% (1.08%$p$) |

---

[18] Target Year means every single year between 1991 and 2009 for DBLP, APS, and KISTI, and between 1996 and 2009 for MEDLINE. Preceding Years mean 5 years prior to each target year in each dataset.

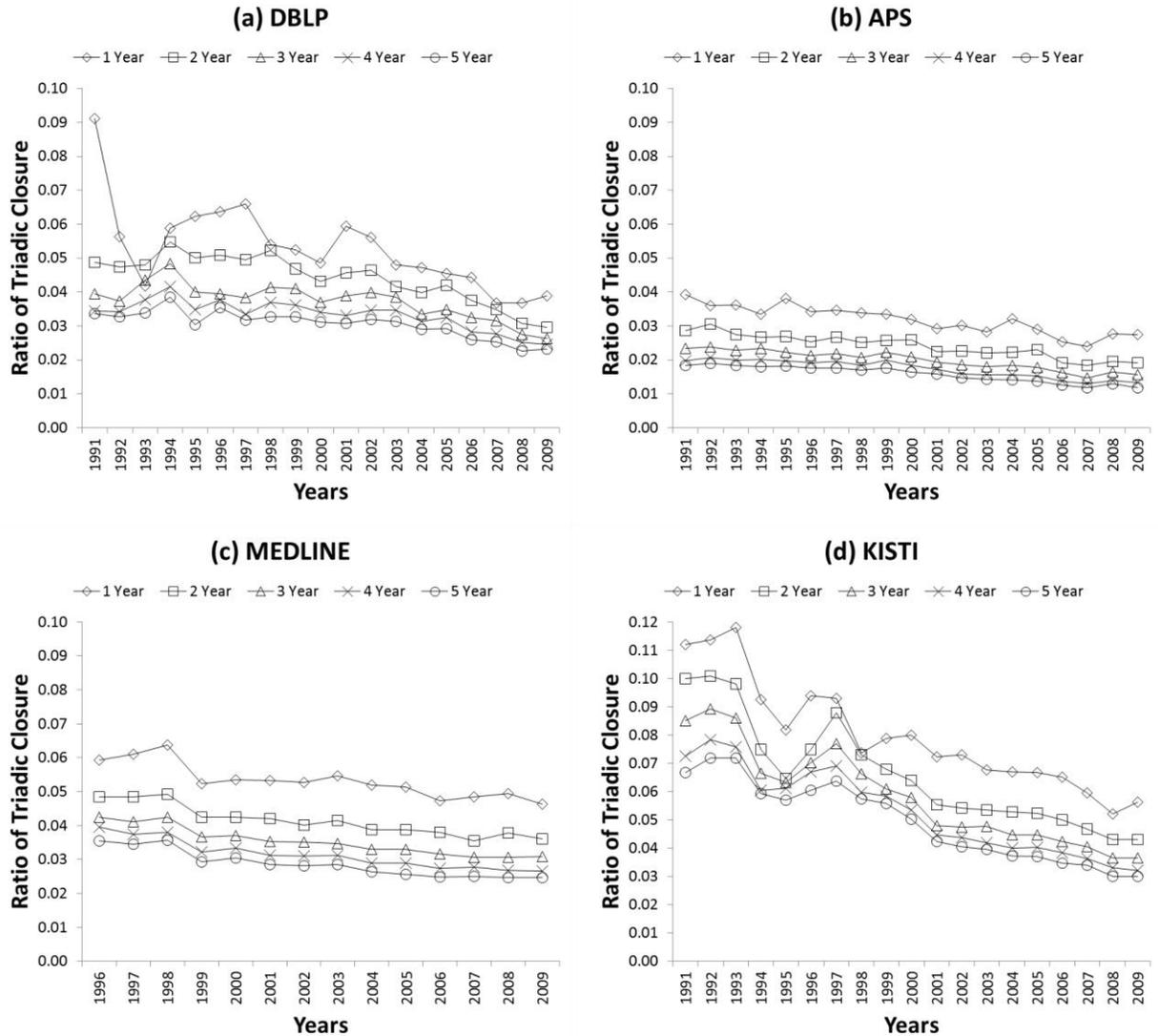

*Figure 2: Ratio of Triadic Closure by TCC with Varying Number of Preceding Years. Preceding years vary from one to five years, while target years are set to one. The larger the amount of preceding years is, the lower are the TCC values. This is because, as the amount of preceding years becomes larger, more 4-paths are likely to happen, leading to the increase of the denominator in TCC calculation. Even with the varied numbers of preceding years, the ration of triadic closure in four datasets when measured with TCC are consistently considerably lower than those estimated by NCC and OCC in Figure 1.*

Next, we analyze how often open 4-paths get closed with the shared coauthor(s) involved. Table 8 illustrates two possible cases of triadic closure: an open 4-paths is closed by the collaboration Y and Z (1) without the shared coauthor X in Paper C, or (2) with the shared coauthor X's involvement in Paper D. Figure 3 shows the ratio (%) of triadic closure where the shared coauthors also participate in a collaboration that completes triadic closure (going back to our general approach of using one target year and five preceding years for TCC calculation). According to our results, less than 50% of shared coauthors in open 4-paths tend to participate in collaborations that close triads.

Table 8: Visualized Concept of Triadic Closure with or without the Shared Coauthor. In CASE1 and CASE2, authors Y and Z who collaborated with X on different papers, come together to work on paper C. CASE1 illustrates a case where X is not involved in collaboration for paper C, while, in CASE2, author X also participates in paper C with Y and Z.

| Case | Network Visualization |
|---|---|
| CASE 1<br>Paper A: author X and author Y<br>Paper B: author X and author Z<br>Paper C: author Y and author Z | |
| CASE 2<br>Paper A: author X and author Y<br>Paper B: author X and author Z<br>Paper D: author X, author Y, and author Z | |

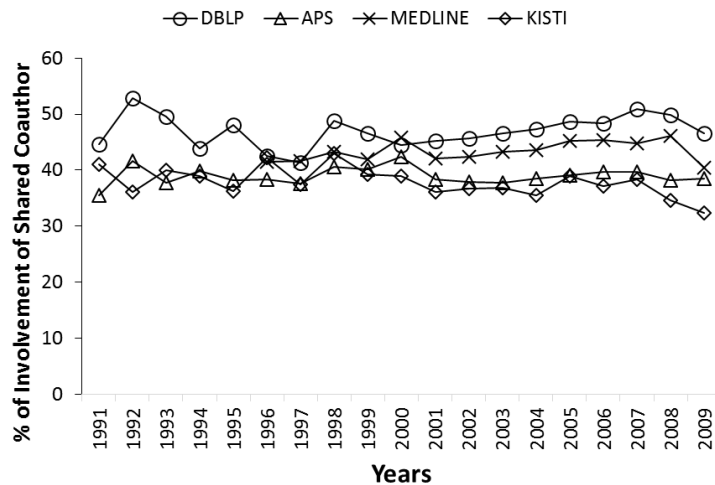

Figure 3: Percentage of Shared Coauthor's Involvement in Collaborations that Close Open 4-paths. Less than 50% of shared coauthors in open 4-paths (five preceding years) tend to participate in collaborations that close triads (in one target year).

Another relevant characteristic of the effects of triadic closure is the number of shared coauthors. Several studies have confirmed that higher numbers of shared coauthors are associated with a stronger likelihood of two yet unconnected authors to collaborate in the future (e.g., Martin et al., 2013; Newman, 2001a). Figure 4 illustrates how often triadic closure happens for all node pairs embedded in open 4-paths that share $n$ third collaborators (target year = 2009 and preceding years = 2004-2008). Similar to the findings from previous studies, as the number of shared collaborators increases, the ratio of node pairs embedded in closed 4-paths also increases in each dataset. For values of 10 and lower on the *x* axis, we see an almost linear increase of triadic closure ratios. However, these ratios are very low even for the case of 5 shared coauthors: only 10% of open 4-paths are closed in every dataset. Even for 10 to 40 shared coauthors, some open 4-paths are not closed.

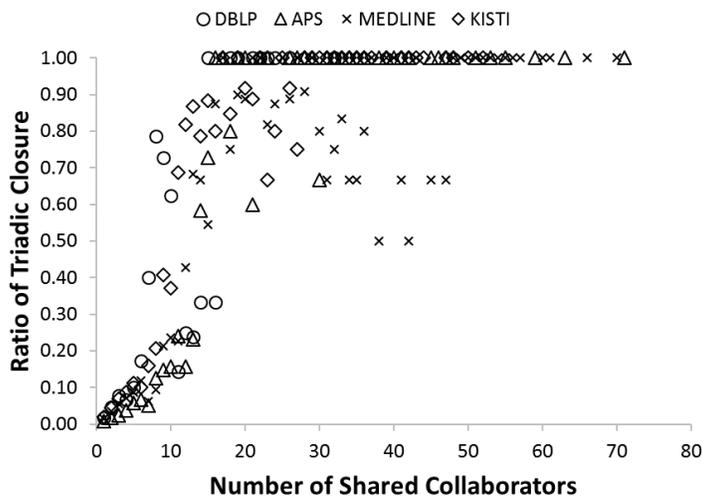

*Figure 4: Ratio of Triadic Closure by TCC as a Function of Number of Shared Collaborators. For illustration, a case of target year = 2009 and preceding years = 2004-2008 is shown. As the number of shared collaborators increases, the ratio of node pairs embedded in closed 4-paths increases in each dataset. Note that even with 10 shared coauthors, approximately 20% of open 4-paths are closed in four datasets.*

4. Conclusion and Discussion

This paper compares triadic closure rates based on three different metrics of closure, namely Newman's measure of closed 2-paths in one-mode networks (NCC), Opsahl's measure of closed 4-paths in two-mode networks (OCC), and an over-time version of Opsahl's measure (TCC). These metrics were applied to four large-scale coauthorship networks. We find that different operationalizations of triadic closure can lead to vastly different interpretations of underlying effects in networks: according to the Newman (2001b) clustering coefficient, for DBLP data, a pair of scholars who have not yet published a paper together but share a common coauthor have a future co-publishing chance of 42.9%, and using the Opsahl (2013) measure (though Opsahl never used his clustering coefficient this way), the chance would be 15.2% on average. If we simply log what truly happens in the data using the over-time measurement proposed in this paper, this likelihood further drops to 3.1% on average. Our findings do not imply that any one measure is right or wrong. If we assume a static structure in which three nodes (e.g., A, B, and C) are fully connected when two of them (e.g., B and C) are both linked to a shared third (e.g., A), then Newman's measure is appropriate. If, however, clustering gets inflated due to network projection, the metric we introduce in this paper allows for staying truthful to effects in data and offers a reliable alternative to the NCC and OCC. If we want to infer edge formation in evolving collaboration networks, OCC may not suffice, while TCC takes temporal effects into account.

The main take away from this study is that triadic closure is rare in scientific coauthoring if we define triadic closure as two authors beginning to collaborate after having had at least one shared coauthor. This finding contradicts several previous studies (starting with seminal work by Rapoport, 1953), and has several implications for future research. In prior work, two disconnected nodes that were both linked to the same third node were assumed to form a tie l. If that is not the case for scientific coauthorship networks, other mechanisms and underlying reasons for link creation or avoidance thereof might apply, such as the creating and maintenance of structural holes (Burt, 2005): sometimes scholars might want to maintain their competitive advantage by brokering between separated others without connecting them (i.e., preventing open paths from closing) rather than introducing people to each other (i.e., facilitating open paths to close). More research is needed in order to understand why only a small percentage of scholars start to co-publish once they share multiple coauthors. For example, there might be organizational or environmental constraints, or other disincentives to forming edges.

This study is not without limitations. First, only open paths that fall into the considered time frames are examined; more long-term (e.g., more than five year horizons) or short-term (within the same year) opportunities and their realizations are not captured. These limitations may partially explain why the OCC and TCC show a gap in estimating triadic closure. The OCC considers all open 4-paths and may detect triadic closure that occurs within a year. By varying window sizes of preceding years, we showed that the overall trends of triadic closure are detected quite consistently, but this does not mean that the TCC correctly captures all instances of closure. Other limiting factors include missing data, e.g., when paths being closed via papers in fields or journals that are not considered in our data. Furthermore, the accuracy of the used datasets might affect the calculations. For example, name disambiguation in the APS dataset is relatively low, which means that merging and splitting of author identities may happen and thus true triadic closure might not be correctly identified (or false triadic closure might be wrongfully counted). Finally, reverse time effects might still happen when people jointly submit a paper to a journal with a lengthy review cycle, and then close a triad in their next conference proceedings publication, which has a shorter review cycle and hence appears prior to the journal paper.